\documentclass{aa}  

\usepackage{graphicx}
\usepackage{caption}
\usepackage{subcaption}
\usepackage{comment}
\usepackage{lscape}
\usepackage{txfonts}
\usepackage[bookmarks=false,colorlinks=true,linkcolor=black,citecolor=cyan,filecolor=black,urlcolor=cyan]{hyperref}

\usepackage{xcolor}

\begin{document}

   \title{On the baryon budget in the X-ray-emitting circumgalactic medium of Milky Way-mass galaxies}

   \author{Yi Zhang \inst{1}\fnmsep\thanks{yizhang@mpe.mpg.de} 
   \and Soumya Shreeram\inst{1}
   \and Gabriele Ponti\inst{2,3,1} 
    \and Johan Comparat\inst{1} 
   \and Andrea Merloni\inst{1}
   \and Zhijie Qu\inst{4}
   \and Jiangtao Li \inst{5}
   \and Joel N. Bregman \inst{6}
   \and Taotao Fang \inst{7}
          }
          
   \institute{Max-Planck-Institut für extraterrestrische Physik (MPE), Gießenbachstraße 1, D-85748 Garching bei München, Germany
   \and
    INAF-Osservatorio Astronomico di Brera, Via E. Bianchi 46, I-23807 Merate (LC), Italy 
    \and Como Lake Center for Astrophysics (CLAP), DiSAT, Università degli Studi dell'Insubria, via Valleggio 11, 22100 Como, Italy
    \and Department of Astronomy, Tsinghua University, Beĳing 100084, People’s Republic of China
    \and Purple Mountain Observatory, Chinese Academy of Sciences, 10 Yuanhua Road, Nanjing 210023, People’s Republic of China
    \and Department of Astronomy, University of Michigan, Ann Arbor, MI 48109, USA
    \and Department of Astronomy, Xiamen University, Xiamen, Fujian 361005, China
   }

   \date{Received ; accepted }

  \abstract
{Recent observations with SRG/eROSITA have revealed the average X-ray surface brightness profile of the X-ray-emitting circumgalactic medium (CGM) around Milky Way (MW)-mass galaxies, offering valuable insights into the baryon budget in these systems. However, the estimation of the baryon mass depends critically on several assumptions regarding the gas density profile, temperature, metallicity, and the underlying halo mass distribution. 
Here, we assess how these assumptions affect the inferred baryon mass of the X-ray-emitting CGM in MW-mass galaxies, based on the stacked eROSITA signal. We find that variations in temperature profiles and uncertainties in the halo mass introduce the dominant sources of uncertainty, resulting in X-ray-emitting baryon mass estimates that vary by nearly a factor of four ($0.8$–$3.5\times10^{11}\,\rm M_\odot$). Assumptions about metallicity contribute an additional uncertainty of approximately $50\%$. 
We emphasize that accurate X-ray spectral constraints on gas temperature and metallicity, along with careful modeling of halo mass uncertainty, are essential for accurately estimating the baryon mass for MW-mass galaxies. Future X-ray microcalorimeter missions will be crucial for determining the hot CGM properties and closing the baryon census at the MW-mass scale.

}


   \maketitle

\section{Introduction}
In the standard cosmological model, baryonic matter constitutes approximately $0.157\pm0.001$ of the total matter content of the Universe and forms the objects we can observe \citep{Planck2020}. A large fraction of these baryons resides in the intergalactic medium (IGM), traced through Ly$\alpha$ forest, O VI and O VII absorbers, and fast radio burst (FRB) dispersion measures; a small fraction of the baryons is found within the potential wells of galaxy systems, in the form of stars and gas \citep[e.g.,][]{Shull2012,Nicastro2018,Connor2024,ChenHW2024}. 
Notably, only the baryon mass ($M_{\rm b}$) enclosed within $R_{\rm 500c}$ of the most massive galaxy clusters ($M_{\rm 500c}\approx10^{15}M_\odot$) matches the cosmological value\footnote{$R_{\rm 500c}$ is the radius at which the mean enclosed density is 500 times the critical density of the Universe. $M_{\rm 500c}$ is the total mass within $R_{\rm 500c}$.} \citep[i.e., $M_{\rm b}/M_{\rm 500c}\approx0.157$,][]{EckertGaspariGastaldello_2021Univ....7..142E}. However, for virialized galaxy structures of decreasing halo mass, an increasing fraction of baryons within $R_{\rm 500c}$ remains unaccounted for \citep{McGaugh2010,Dai2012,Li2017,EckertGaspariGastaldello_2021Univ....7..142E,Dev2024,Popessofb2024}.
There are two main explanations for the low detected baryon mass in lower-mass galaxy systems.
First, baryons may have been ejected beyond $R_{\rm 500c}$ or even depleted from the gravitational potential wells of galaxies via active galactic nucleus (AGN) feedback. This interpretation is supported by the observed flatter gas density profiles of the intracluster medium (ICM) compared to those of dark matter halos \citep[e.g.,][]{Sun2009,Eckert2016}, as well as by numerical simulations \citep{Gaspari2012,Sorini2022,Wright2024,Nelson2024,Ayromlou2024}.
Second, the low observed baryon fraction is due to observational limitations that the CGM (or the intra-group medium) of galaxies with $M_{\rm 500c}<10^{14}M_\odot$ has low X-ray emissivity, making it difficult to detect.

The mass of the MW’s X-ray-emitting CGM has been estimated from O VII and O VIII emission and absorption line measurements, based on tens to hundreds of {\it XMM-Newton} and {\it Chandra} sightlines \citep{Gupta2012, Miller2013, Henley2013, Miller2015,Fang2015,Hodges-Kluck2016,Nicastro2016,LiYY2017,Faerman2017,Bregman2018}. 
The reported baryon mass estimates are summarized in Appendix~\ref{Sec_MWb}. In general, the X-ray-emitting gas mass within $250\,\rm kpc$ ranges from less than $3\times10^{10}\,M_\odot$ to values more than $10^{11}\,M_\odot$.
One reason for this wide range is our location: the gas density profile slope is difficult to constrain due to projection effects. Indeed, several studies report that a disk-like hot component with a scale height of $<3\,\rm kpc$ and a temperature of $T\approx0.2$ keV can explain most of the observed X-ray emission \citep{Yao2007,Kaaret2020, Locatelli2023}.

In nearby massive galaxies ($M_*\sim10^{11}\,M_\odot$) with extended X-ray emission detected out to 10--100 kpc, the gas density profile of the X-ray-emitting CGM is typically modeled with a beta profile, with $\beta\approx 0.4-0.6$ \citep[][see also the review by \citealt{Bregman2018}]{Yao2009,Wang2010,Humphrey2011,Anderson2011, Dai2012,Bogdan2013, Anderson2016, Li2018, Bregman2022}. The total baryon content within the virial radius requires extrapolating the gas density profile, which introduces significant uncertainty.

Apart from pointed observations toward individual galaxies, the average X-ray surface brightness ($S_{\rm X}$) in the $0.5-2$ keV energy band of MW-mass galaxies has been measured through stacking analyses out to $\approx 150\,\rm kpc$ \citep{Zhang2024profile}. While targeted X-ray observations of the MW and nearby galaxies provide constraints on the temperature of the X-ray-emitting CGM, determining the temperature from stacked measurements across the broad 0.5--2 keV energy band remains challenging \citep{Toptun2025}. This difficulty arises from the limited photon statistics available for X-ray spectral analysis.
In light of these limitations, we discuss in this paper how different assumptions about the temperature, metallicity, gas density profile, and the uncertainty in the underlying galaxy halo mass affect the estimation of the X-ray-emitting gas mass within the virial radius ($R_{\rm vir}$) of MW-mass galaxies. 
We focus on the MW-mass galaxies, as they dominate the stellar mass budget at low redshift \citep{Driver2022}, and they are the most sensitive to the various assumptions\footnote{We discuss the baryon mass uncertainty for more massive galaxies in Appendix~\ref{Sec_moremassive}.}.
For our analysis, we adopt the $S_{\rm X}$ profile of the MW-mass central galaxy (CEN) sample from \citet{Zhang2024profile} as a benchmark to evaluate the associated X-ray-emitting gas mass (or baryon fraction) with different assumptions. We assume an NFW-shaped dark matter halo for MW-mass galaxies \citep{NFW1997}, with a fiducial virial mass (radius) of $M_{\rm vir}=1.3\times10^{12}\,M_\odot$ ($R_{\rm vir}=285\,\rm kpc$). 

\section{X-ray emission and baryon mass}

The integrated 0.5--2 keV X-ray emission produced by fully ionized plasma with temperature ($T$), metallicity ($Z$), and hydrogen number densities ($n_{\rm H}$) is given by:  
\begin{equation}
\epsilon_{\rm X,\,0.5-2\,keV}=\Lambda_{\rm 0.5-2\,keV} (T,Z) \mu_{\rm e} n_{\rm H}^2 \,{\rm [erg\ s^{-1}\ cm^{-3}]},
\end{equation}
where $\Lambda_{\rm 0.5-2\,keV}(T,Z)$ is the X-ray cooling function in the 0.5--2 keV band, which depends on the plasma temperature and metallicity. $\mu_{\rm e}\approx 1.16$ is the ratio of electron to hydrogen number densities \citep[][]{Lodders2003}.

We assume that the hot CGM is spherically symmetric and volume-filling\footnote{The volume-filling factor ($f$) of hot CGM is heavily uncertain. The inferred baryon mass scales as $f^{-1/2}$.}, such that $T$, $Z$, and $n_{\rm H}$ depend only on radius $r$. The X-ray surface brightness profile $S_{\rm X}(r)$ is obtained by integrating $\epsilon_{\rm X}$ along the line of sight ($dl$) as:
\begin{equation}
S_{\rm X}(r)=\int \epsilon_{\rm X}(r) \,dl\,{\rm [erg\ s^{-1}\ kpc^{-2}]}.
\end{equation}
Since $S_{\rm X}$ scales with $n_{\rm H}^2$, the X-ray emission is dominated by the densest regions (e.g., the inner CGM).
Assuming the plasma is in collisional ionization equilibrium, $\Lambda_{\rm 0.5-2\,keV}$ is highly sensitive to both the temperature and metallicity of the plasma, particularly when the plasma temperature is around 0.1 keV \citep[][also see Fig.6 in \citealt{Grayson2025}]{Bohringer2010}.
For plasma with $T\approx0.1-0.2$ keV, the 0.5--2 keV spectrum is dominated by lines from ions such as O VII, O VIII, Fe XVII. These lines exhibit a strong temperature dependence, as the corresponding ionization fractions of the ions decrease sharply below $\sim$0.2 keV.
For example, when the temperature decreases from 0.1 keV to 0.09 keV, $\Lambda_{\rm 0.5-2\,keV}$ drops by approximately 50\% (assuming metallicity of 0.3$\,Z_\odot$ and fixed H and He abundances).
Similarly, if the metallicity decreases from 0.3$\,Z_\odot$ to 0.1$\,Z_\odot$ (assuming $T=0.1$ keV), $\Lambda_{\rm 0.5-2\,keV}$ decreases by about 60\%. Therefore, deriving the gas mass from $S_{\rm X}$ profile of MW-mass galaxies (such as \citealt{Zhang2024profile}), whose CGM is expected to have a temperature on the order of $0.1\,\rm keV$, requires careful assumptions about $T(r)$, $Z(r)$, and $n_{\rm H}(r)$ profile.

The X-ray emission observed around nearby galaxies suggests that the hot gas density profile can be described by a beta model \citep[e.g.,][]{Li2013a}:
\begin{equation}\label{Eq_beta}
n_{\rm H}=n_{0}(1+(r/r_{\rm c})^2)^{-3\beta/2}\,[\rm cm^{-3}],
\end{equation}
where $n_{0}$ is the central gas density, $r_{\rm c}$ is the core radius, and $\beta$ determines the slope of the $n_{\rm H}(r)$ profile.
The mass of the X-ray-emitting gas within the virial radius, $R_{\rm vir}$, is then calculated as
\begin{equation}
M_{\rm X-ray-emitting, <R_{\rm vir}}=4\pi \mu_{\rm H} m_{\rm p}\int_0^{R_{\rm vir}} n_{\rm H}r^2 dr\,[M_\odot],
\end{equation}
where $\mu_{\rm H} m_{\rm p}$ is the mean mass per hydrogen ($\mu_{\rm H}\approx 1.32$).
A typical range of $\beta=0.4$–$0.6$ within $\approx$100 kpc of galaxies with $\log(M_*/M_\odot)=10.5$–$11.5$ implies a radial baryon mass profile of $dM_{\rm X-ray-emitting, <R_{\rm vir}}/dr \propto r^{0.2}$–$r^{0.8}$ \citep[see review in][]{Bregman2018}. This indicates that the baryon mass is dominated by the outer halo\footnote{We note that $\beta>1$ at large radii is required to assure the CGM mass is converged. We do not consider the convergence of the CGM mass in our fiducial assumption, instead, the assumed modified Vikhlinin profile (see Sect.~\ref{Sec_nh}) satisfies the need of CGM mass convergence.}, in contrast to the X-ray emission, which is dominated by the inner halo, underscoring the importance of carefully assessing the assumptions used to derive the gas density profile from the $0.5-2$ keV $S_{\rm X}$ profile.

We begin by adopting a fiducial model in which the gas density profile follows a beta model with $\beta=0.4$ and $r_{\rm c}=5\,\rm kpc$ \citep{Zhang2024profile}, a metallicity of $0.3\,Z_\odot$ \citep{Anderson2010}, and an isothermal temperature of $T=0.12\,\rm keV$ (see discussion in Sect~\ref{Sec_T}). We also take the fiducial $M_{\rm vir}$ and $R_{\rm vir}$. By fitting this model to the observed $S_{\rm X}$ profile from \citet{Zhang2024profile}, we derive an X-ray-emitting gas mass within $R_{\rm vir}$ of $M_{\rm X-ray-emitting, <R_{\rm vir}}=1.26\pm0.09\times10^{11}M_\odot$ and a corresponding baryon fraction of $f_{\rm X-ray-emitting, <R_{\rm vir}}=M_{\rm X-ray-emitting, <R_{\rm vir}}/M_{\rm vir}=0.097\pm0.007$ with a reduced $\chi^2=0.3$ for the fit\footnote{We use MCMC to perform the fit; 0.097 is the median value and 0.007 is the 1$\sigma$ uncertainty.}.
Note that $f_{\rm X-ray-emitting, <R_{\rm vir}}$ does not include the baryon content in stars (the median stellar mass of stacked galaxies in \citet{Zhang2024profile} is $M_*=5.5\times10^{10}M_\odot$, which yields $f_{\rm b, *}=M_*/M_{\rm vir}= 0.042$), or the cooler, non-X-ray-emitting phase of the CGM \citep{Werk2014,Peeples2014,TumlinsonPeeplesWerk_2017ARAA..55..389T,ChenHW2020, Tchernyshyov2022,Ng2025}. A more detailed discussion is provided in Appendix~\ref{Sec_MWb}.

\begin{figure}[ht]
\centering
\includegraphics[width=0.99\columnwidth]{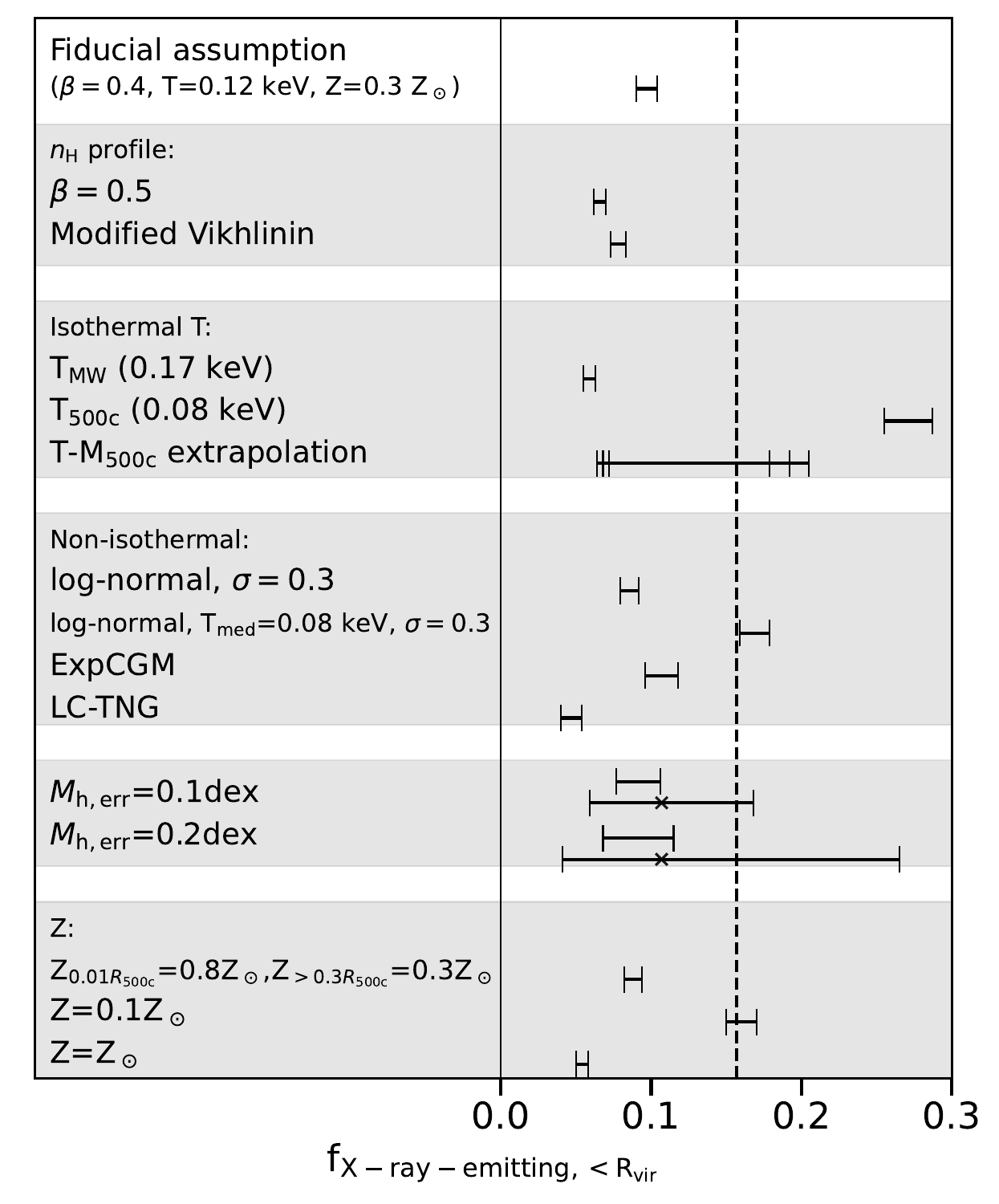}
\caption{Derived baryon mass fraction of the X-ray-emitting gas within $R_{\rm vir}$ based on the mean X-ray surface brightness profile of MW-mass central galaxies from \citet{Zhang2024profile}. We evaluate the effects of different assumptions on the inferred baryon fraction, including: $n_{\rm H}$ profile (Sect.~\ref{Sec_nh}); isothermal temperature value (Sect.~\ref{Sec_T}); log-normal distribution of temperature, ExpCGM model, and TNG-based forward model (Sect.~\ref{Sec_Tlog}); halo mass uncertainty (Sect.~\ref{Sec_Mh}); and metallicity (Sect.~\ref{Sec_Z}). The vertical dashed line indicates the cosmological baryon fraction ($f_{\rm b}=$0.157).}
\label{Fig_fb}
\end{figure}

In the following sections, we vary the assumed $T$, $Z$, and $n_{\rm H}$ profile relative to the fiducial case and consider the effect of uncertainties in the halo mass of MW-mass galaxies. The results are summarized in Fig.~\ref{Fig_fb}.

\subsection{Dependence on the gas density profile}\label{Sec_nh}

The $S_{\rm X}$ profile of MW-mass galaxies beyond $\sim$150 kpc has not yet been measured \citep{Zhang2024profile}, leading to significant uncertainties in the $n_{\rm H}$ profile at large radii. In this section, we examine how modifications to the assumed $n_{\rm H}$ profile affect the inferred $M_{\rm X-ray-emitting, <R_{\rm vir}}$ and $f_{\rm X-ray-emitting, <R_{\rm vir}}$.

To explore the possible variations in the CGM gas density profile, we take guidance from the better-measured $S_{\rm X}$ profile of the ICM.
Typical ICM $S_{\rm X}$ profile exhibits complicated structures, including a central core, and a steep decline at large radii \citep{Vikhlinin2006,Arnaud2010,Eckert2016,Popesso2024_profile}. As a result, a single beta model is insufficient to describe the ICM, and the Vikhlinin model is commonly adopted \citep{Vikhlinin2006}.
We modify the fiducial $n_{\rm H}$ profile to (i) a steeper beta model with $\beta=0.5$, and (ii) a modified Vikhlinin profile, 
\begin{equation}\label{Eq_Vikh}
    n_{\rm H}=n_{0}(1+(r/r_{c})^2)^{-3\beta/2}(1+(r/r_{s})^3)^{-\epsilon/6},
\end{equation}
with the fiducial values of $\beta$ and $r_{\rm c}$, and adopting $r_{\rm s}=200$ kpc and $\epsilon=4$, as suggested by \citet{Shreeram2025I}\footnote{$n_{\rm H}$ profile with $\beta=0.6$ ($\beta=0.3$) would provide poor fit to the observed $S_{\rm X}$ profile with reduced $\chi ^2=3.2\,(2.3)$.}. Fitting the $\beta=0.5$ model to the observed $S_{\rm X}$ profile yields $M_{\rm X-ray-emitting, <R_{\rm vir}}=8.6\pm0.5\times10^{10}M_\odot$ ($f_{\rm X-ray-emitting, <R_{\rm vir}}=0.066\pm0.004$, reduced $\chi^2=1.2$). Fitting the modified Vikhlinin model yields $M_{\rm X-ray-emitting, <R_{\rm vir}}=10.1\pm0.7\times10^{10}M_\odot$ ($f_{\rm X-ray-emitting, <R_{\rm vir}}=0.078\pm0.005$, reduced $\chi^2=0.8$). The $f_{\rm X-ray-emitting, <R_{\rm vir}}$ values inferred from these two $n_{\rm H}$ models are about 30\% and 20\% lower, respectively, than the fiducial value. Larger ($>2$ times) optical galaxy samples from large-volume surveys, such as DESI BGS and 4MOST, will become available within the next five years, substantially improving the measurements of the $S_{\rm X}$ profile at large radii \citep{HahnWilsonRuiz-Macias_2023AJ....165..253H, Driver2019_WAVES,FinoguenovMerloniComparat_2019Msngr.175...39F}. Ultimately, narrow-band (e.g., in O VII and O VIII lines) imaging enabled by future microcalorimeter X-ray telescopes will suppress the cosmic X-ray background, enhance the signal-to-noise ratio of the hot CGM X-ray emission, and constrain the gas density profile at large radii with high precision \citep{LJT2020,Barret2023,hubs2023}.

\subsection{Temperature dependence}\label{Sec_T}

The temperature of the MW’s X-ray-emitting halo has been measured as $0.19\pm0.05\,\rm keV$ using 110 {\it XMM-Newton} sightlines \citep{Henley2013}, $0.166 \pm 0.005\,\rm keV$ from the {\it HaloSat} all-sky survey \citep{Bluem2022}, $0.176 \pm 0.008\,\rm keV$ from {\it Suzaku} observations \citep{Gupta2021}, and $0.153-0.178\,\rm keV$ in the eFEDS field \citep{Ponti2023}. These temperatures are derived from line-of-sight-integrated emission or absorption, and the radial temperature variation remains unknown. Temperatures inferred from X-ray emission are biased towards the densest regions (e.g., those near the Galactic disk), which may not represent the conditions in lower-density regions of the halo.

For nearby star-forming galaxies with stellar masses $1$–$10\times10^{10}\,M_\odot$, the X-ray-emitting gas temperature ranges from $0.2$ to $0.8\,\rm keV$ \citep[][]{Li2017}. For early-type galaxies with $M_*\approx 10^{11}\,M_\odot$, the temperatures measured within 30 kpc are typically $1.4-2$ times higher than the corresponding virial temperatures \citep{Goulding2016, Bregman2018}. The relatively higher temperatures within $\sim$ 50 kpc suggest that stellar feedback may substantially heat the gas, and that selection biases toward X-ray-luminous or hotter systems may be present \citep{Li2017}, as also found at the galaxy group scale \citep{Marini2025}.

Overall, the temperature profile of MW-mass galaxies remains poorly constrained. Varying the temperature in the fiducial model to 0.17 keV (i.e., consistent with the MW's CGM) yields an X-ray-emitting gas mass of $7.7\pm0.5\times10^{10}M_\odot$ and a corresponding baryon fraction of $0.059\pm0.004$, slightly higher than the MW estimates ($0.027-0.047$, \citet{Miller2015}).

The conventional theoretical assumption for the CGM temperature is the isothermal `virial' temperature ($T_{\phi}$), which is related to the circular velocity $V_{\rm c}$ of the dark matter (DM) halo by \citep{Maller2004}:
\begin{equation}
k_{\rm B}T_{\phi}=\frac{\mu m_{\rm p}}{2} V_{\rm c}^2 = 0.03\,{\rm keV} \left(\frac{V_{\rm c}}{100\,\rm km\,s^{-1}}\right)^2. \label{Eq_Tphi}
\end{equation}
Under the singular isothermal sphere assumption, the DM halo density follows $\rho_{\rm DM}\propto r^{-2}$, implying a constant $V_{\rm c}(r) = \sqrt{GM(r)/r}$. The MW's circular velocity is approximately $200\,\rm km/s$ \citep{Bland2016,Beordo2024}, corresponding to a virial temperature of $\approx 0.12\,\rm keV$, which we adopt as the fiducial value\footnote{The assumptions of an isothermal halo and an NFW-shaped DM profile are not strictly self-consistent. However, since X-ray emission from the CGM is primarily detected at small radii, $V_{\rm c}$ can be approximated as constant.}. 

In reality, for an NFW-shaped halo, $V_{\rm c}$ is not constant. For the MW, $V_{\rm c}$ is about $235\,\rm km/s$ at the solar radius and decreases roughly as $r^{-1/2}$ \citep{Jiao2023,Beordo2024}. If we instead assume an isothermal halo with $T_{\phi}$ evaluated at $R_{\rm 500c}$ (about $0.08\,\rm keV$) we obtain an X-ray-emitting gas mass of $3.5\pm0.2\times10^{11}M_\odot$ and $f_{\rm X-ray-emitting, <R_{\rm vir}}=0.271\pm0.016$, which significantly exceeds the cosmic baryon fraction.

We note that \citet{Zhang2024profile} estimated the hot baryon mass using the isothermal assumption with $T_{\phi}$ taken at $R_{\rm vir}$ ($T_{\rm vir}$). As discussed above, and further in later sections, the isothermal $T_{\rm vir}$ assumption is oversimplified, and the associated uncertainty in the derived baryon fraction is significantly underestimated. Moreover, the $T_{\rm vir}$ adopted in \citet{Zhang2024profile} is overestimated by a factor of about 3, such that $T_{\rm vir}$ of MW-mass galaxies was taken as $0.22\,\rm keV$, while the correct $T_{\rm vir}$ should be $0.06\,\rm keV$. By coincidence, the overestimated $T_{\rm vir}=0.22\,\rm keV$ is close to the observed temperature of the MW halo. 
In Appendix~\ref{Sec_moremassive}, we discuss how $f_{\rm X-ray-emitting, <R_{\rm vir}}$ for other mass bins in \citet{Zhang2024profile} depends on different assumptions regarding temperature, metallicity, and halo mass.
However, a more accurate determination of $f_{\rm X-ray-emitting, <R_{\rm vir}}$ requires a forward-modeling approach that self-consistently accounts for the stacking procedure (see Sect.~\ref{Sec_other}), as well as future X-ray spectral analyses.

Because $M_{\rm 500c}\propto R_{\rm 500c}^3$, Eq.~\ref{Eq_Tphi} suggests $k_{\rm B}T_{\phi} \propto M_{\rm 500c}^{2/3}$, the so-called self-similar model, which assumes that the thermodynamic properties of the ICM are determined solely by gravity and that the gas is in hydrostatic equilibrium \citep{Kaiser1986,Lovisari2021}. 
Observations and simulations, however, indicate a shallower scaling, with $k_{\rm B}T$–$M_{\rm 500c}$ slopes of $0.57$–$0.66$ for galaxy groups and clusters \citep{Sun2009,Eckmiller2011,Lovisari2015,Kettula2015,Pop2022,Toptun2025}, likely due to additional heating from feedback processes. Using $T=4.3\,\rm keV$ at $M_{\rm 500c}=3\times10^{14}M_\odot$ \citep{Lovisari2021} and extrapolating this relation down to MW-mass halos yields temperatures of $0.09$–$0.15\,\rm keV$, corresponding to $M_{\rm X-ray-emitting, <R_{\rm vir}}=2.5\pm0.2\times10^{11}M_\odot$ ($f_{\rm X-ray-emitting, <R_{\rm vir}}=0.192\pm0.013$) and $M_{\rm X-ray-emitting, <R_{\rm vir}}=8.8\pm0.5\times10^{10}M_\odot$ ($f_{\rm X-ray-emitting, <R_{\rm vir}}=0.068\pm0.004$), as shown in Fig.~\ref{Fig_fb} and listed in Table~\ref{Table_fbMs}.

In summary, variations in the assumed isothermal $T$ cause $f_{\rm X-ray-emitting, <R_{\rm vir}}$ to vary dramatically between 6--27\%, underscoring the importance of reliable temperature measurements for accurately estimating the gas mass in MW–mass galaxies. This large uncertainty in $f_{\rm X-ray-emitting, <R_{\rm vir}}$ arises from the strong temperature dependence of $\Lambda_{\rm 0.5-2\,keV}$ on $T$ for plasma with $T\sim0.1\,\rm keV$. For more massive galaxies, where the CGM temperature is typically $T>0.2\,\rm keV$, $\Lambda_{\rm 0.5-2\,keV}$ becomes less sensitive to $T$, for example, increasing $T$ by 10\% (from 0.2 to 0.22 keV) raises $\Lambda_{\rm 0.5-2\,keV}$ by $\sim$13\%. As a result, the inferred X-ray-emitting gas mass of galaxy groups and clusters depends less strongly on the assumed $T$, see the discussion in Appendix~\ref{Sec_moremassive}. 

\subsection{Non-isothermal temperature} \label{Sec_Tlog}

Small-scale temperature fluctuations driven by turbulence have been suggested by both observations \citep{Faerman2017, Qu2024,Lovisari2024} and simulations \citep{Blaisdell1993, Padoan1997, McCourt2012}. These fluctuations are often modeled with a log-normal temperature distribution:
\begin{equation}
f(k_{\rm B}T) = \frac{1}{\sigma k_{\rm B}T \sqrt{2\pi}}\,\exp\left(-\frac{(\ln T - \ln T_{\rm med})^2}{2\sigma^2}\right),
\end{equation}
where $T_{\rm med}$ is the median temperature and $\sigma$ is the logarithmic scatter. A model with $T_{\rm med} = 0.12\,\rm keV$ and $\sigma \approx 0.3$ \citep{Faerman2017}, or a normal distribution with $T_{\rm med} = 0.225\,\rm keV$ and $\sigma_{\rm T} \approx 0.023\,\rm keV$ \citep{Kaaret2020}, has been proposed to reproduce the MW's O VII/O VIII absorption and emission data. 

To quantify the impact of a log-normal temperature distribution on the derived gas mass, we adopt $T_{\rm med}$ equal to the fiducial temperature and $\sigma=0.3$, following \citet{Faerman2017}. We find that the log-normal distribution introduces a high-temperature tail, increasing the mean X-ray emissivity by about 30\%. Consequently, to maintain the same observed $S_{\rm X}$, the inferred $M_{\rm X-ray-emitting, <R_{\rm vir}}$ and $f_{\rm X-ray-emitting, <R_{\rm vir}}$ decrease by about 13\% relative to the fiducial isothermal case.
The influence of this effect can be even more pronounced at lower temperatures. For example, a plasma with a log-normal distribution of $T_{\rm med}=0.08\,\rm keV$, and $\sigma=0.3$ emits X-rays that are $\sim$2.5 times brighter than a plasma with an isothermal $T=0.08\,\rm keV$ (i.e., $T_{\rm 500c}$ that we adopted in Sect.~\ref{Sec_T}). This would yield $f_{\rm X-ray-emitting, <R_{\rm vir}}=0.17\pm0.01$, approximately 160\% lower than the value obtained under the isothermal $T=0.08\,\rm keV$ assumption.

The CGM temperature is also unlikely to remain constant with radius, as suggested by both simulations \citep{Truong2023,Ramesh2023,Shreeram2025III} and observations of the ICM \citep{Vikhlinin2005,Arnaud2010} and the MW's CGM \citep{Yao2007,Yao2009,Feldmann2013, Bregman2018}. Using O VIII/O VII and Fe-L/(O VIII + O VII) line ratios from {\it XMM-Newton} archival data, \citet{Qu2024} found that the halo temperature decreases from $\sim$0.18 keV toward the Galactic center to $\sim$0.07 keV in the anti-Galactic direction. 

Several theoretical models (e.g., the precipitation model and the cooling flow model) have been proposed to explain the structure of the CGM, and all predict a temperature that decreases with radius \citep[see discussion and references in][]{Singh2024}. While a comprehensive comparison of these models is beyond the scope of this paper, we adopt the physically motivated ExpCGM framework\footnote{\url{https://gmvoit.github.io/ExpCGM/}}, which provides a self-consistent description of the gaseous halo using a minimal set of four parameters. We assume purely thermal pressure ($f_{\rm th}=1$), hydrostatic equilibrium ($f_{\rm \phi}=1$, $dP/\rho\,dr = -d\Phi/dr$), an NFW dark matter halo, and a generalized NFW-like pressure shape function ($\alpha(r)$):
\begin{equation}
\alpha(r)=-\frac{d\ln P(r)}{d\ln r} = \left(\alpha_0 + \alpha_1 \frac{(r/r_{\rm max})^{1.1}}{1 + (r/r_{\rm max})^{1.1}}\right),
\end{equation}
where $r_{\rm max}$ is the radius where the circular velocity peaks (held fixed). The equilibrium pressure ($P_{\rm HSE}(r)$) profile is then
\begin{equation}
P_{\rm HSE}(r) = P_0 \exp\left(-\int_0^r\frac{\alpha(r)}{r}dr\right),
\end{equation} 
where $P_0$ is the pressure normalization, which is a function of $f_{\rm b}$.
The equilibrium temperature profile relates to $T_\phi$ and $\alpha$ as
\begin{equation}
T_{\rm HSE}(r) = 2T_\phi(r)|\alpha(r)|^{-1},
\end{equation} 
and the gas density profile is related to $P_{\rm HSE}$ and $T_{\rm HSE}$ as $n_{\rm H}=P_{\rm HSE}(r)/(k_{\rm B}T_{\rm HSE}(r))$.

For example, in the ExpCGM framework, an isothermal condition corresponds to $\alpha(r)=2$, which yields $P_{\rm HSE} \propto r^{-2}$, $T_{\rm HSE} = T_\phi$, independent of the pressure gradient. If $|\alpha|$ is smaller (higher) than $2$, $T_{\rm HSE}$ is higher (lower) than $T_\phi$. 

Using the ExpCGM framework, we derive pressure, temperature, and gas density self-consistently from three free parameters $\alpha_0$, $\alpha_1$, and $f_{\rm b}$, by fitting to the observed $S_{\rm X}$ profile. To ensure gas confinement, we impose $d\ln P/d\ln r > 1.5$ at $R_{\rm vir}$. The best fit to the observed $S_{\rm X}$ profile yields $\alpha_0=0.6\pm0.2$, $\alpha_1=1.1\pm0.3$, and $f_{\rm X-ray-emitting, <R_{\rm vir}} = 0.11 \pm 0.01$ ($M_{\rm gas, <R_{\rm vir}}=1.4\pm0.1\times10^{11}M_\odot$). The inferred temperature decreased from $\sim 0.2\,\rm keV$ at $r=30\,\rm kpc$ to $\sim 0.1\,\rm keV$ at $R_{\rm 500c}$.

However, the assumptions we made under the ExpCGM framework may oversimplify the physical state of the CGM. For MW-mass galaxies, the CGM (at least in the inner halo) may not be in complete hydrostatic equilibrium \citep{Faucher-GiguereOh_2023,Jana2024,Kakoly2025}. Moreover, non-thermal pressure support (e.g., from turbulence and cosmic rays) and cooler CGM phases may contribute significantly to the total pressure budget, as suggested by the simulations \citep{Gaspari2013,Quataert2025,Kakoly2025}. 
Nevertheless, as a self-consistent model, the ExpCGM framework enables simultaneous fitting to multiple observables (e.g., X-ray, Sunyaev–Zel’dovich (SZ) effect, and FRB dispersion measures). Actually, the Descriptive Parametric Model (DPM) introduced by \citet{Oppenheimer2025} follows a similar logic to the ExpCGM, but further accounts for the dependence of halo properties on halo mass and redshift. \citet{Oppenheimer2025} apply the DPM formalism to explain diverse observables, including X-ray surface brightness profiles, Compton $y$ parameter profile from thermal SZ observations, electron dispersion measures, O VI column density profile, etc.

Cosmological simulations are an essential tool for interpreting systematics in the observed $S_{\rm X}$ profile, for example, the projection effect, miscentering, halo mass uncertainties, and mass distribution, which have been discussed in, e.g.,\citet[][]{Shreeram2025I, Popesso2024_sim,Grayson2025}. Simulations with different feedback prescriptions exhibit distinct baryonic behaviors and provide valuable insight into the properties of the hot CGM, as they incorporate a wide range of physical processes \citep{Wright2024, Braspenning2024,Lau2025_camels,Dolag2025}.
The TNG300-based lightcone (LC-TNG) developed by \citet{Shreeram2025I} is specifically designed to model the hot gas emission in the redshift range $0.03 \lesssim z \lesssim 0.3$. LC-TNG generates self-consistent mock X-ray observations using the intrinsic gas cell information from TNG300. Halos are projected onto the sky, and $S_{\rm X}$ profiles are extracted following the same stacking strategy used in observations, allowing direct comparison with real data \citep{Shreeram2025II}. We find that the X-ray profile of halos with $M_{\rm 200c}\approx2\times10^{12}\,M_\odot$ in LC-TNG agrees well with observations within 80 kpc\footnote{Beyond $\sim$80 kpc, the LC-TNG $S_{\rm X}$ profile appears steeper than observed. This discrepancy can be mitigated by forward-modeling the halo mass distribution of the stacked sample, as in \citet{Shreeram2025II}.}, yielding $f_{\rm X-ray-emitting, <R_{\rm vir}}=0.0469\pm0.007$ ($M_{\rm X-ray-emitting, <R_{\rm vir}}=6.1\pm0.9\times10^{10}M_\odot$). 
The simulated temperature profile decreases with radius, from $\sim$0.2 keV at the center to $\sim$0.1 keV at $R_{\rm 500c}$ (see Fig. 5 in \citealt{Shreeram2025III}). 
It is important to note that simulations typically compute mass-weighted or volume-weighted temperatures of hot (i.e., $T>10^{5.5}\,\rm K$) gas particles, whereas observations measure an emission-weighted temperature, $T_{\rm emi}$, via spectral fitting. \citet{Pop2022} find that $T_{\rm emi}$ is typically $\sim$9\% higher than the mass-weighted temperature and exhibits about twice the scatter. This systematic difference should be considered when comparing simulated and observed temperature measurements.

Future X-ray microcalorimeter telescopes with spectral resolution of $\sim 2\,\rm eV$ can separate prominent O VII and O VIII lines associated with the hot CGM around galaxies at $z>0.01$ from MW foreground emission, and constrain the multi-temperature components of the hot CGM through detailed X-ray spectral analysis \citep{ZhangYN2022}, along with a large effective area or field of view to ensure sufficient photon collection within reasonable exposure times \citep[e.g.,][]{Cui2020,Cruise2025}.

\subsection{Dependence on the halo mass}\label{Sec_Mh}

The MW's halo mass is inferred through tracers such as halo stars, globular clusters, satellite galaxies, and tidal streams \citep{Gibbons2014,Vasiliev2019, Callingham2019,Cautun2020, Shen2022,Ibata2024}, which yield a broad range of estimates, from $5.5 \times 10^{11}\,M_\odot$ to $2.6 \times 10^{12}\,M_\odot$, with a commonly adopted value of $1.3 \pm 0.3 \times 10^{12},M_\odot$ \citep[namely about 0.1 dex uncertainty][]{Bland2016,McMillan2017}.
For external galaxies, halo masses are typically estimated from the circular velocity function, satellite kinematics, and abundance matching \citep{HYPERLEDA20031, CrookHuchraMartimbeau_2007ApJ...655..790C,Yang2007, TempelKipperTamm_2016AA...588A..14T, MandelbaumWangZu_2016MNRAS.457.3200M, Tinker2021,Tinker2022,Popesso2024_profile, Marini2025}, with typical uncertainties of $\lesssim \,0.2 \,\rm dex$ \citep{Tinker2022}. Notice that the uncertainty of the mean halo mass of a complete galaxy sample is much smaller than $0.2 \,\rm dex$, instead, the scatter of the halo mass of the galaxy sample is about $0.15-0.2$ dex \citep{Matthee2017,Tinker2017,Popesso2024_profile}.

We assess the effect of halo mass uncertainty by assuming 0.1 dex (representative of the MW) and 0.2 dex. The halo mass uncertainty affects the estimation of $f_{\rm X-ray-emitting, <R_{\rm vir}}$ in two ways. First, it has a direct effect as the denominator when calculating $f_{\rm X-ray-emitting, <R_{\rm vir}}$. Second, it has a side effect if the temperature is assumed from the $k_{\rm B}T$–$M_{\rm 500c}$ relation. Considering only the direct effect, uncertainties of 0.1 dex and 0.2 dex yield $M_{\rm X-ray-emitting, <R_{\rm vir}}=1.2\pm0.2\times10^{11}M_\odot$ ($f_{\rm X-ray-emitting, <R_{\rm vir}} = 0.09 \pm 0.01$) and $M_{\rm X-ray-emitting, <R_{\rm vir}}=1.2\pm0.3\times10^{11}M_\odot$ ($f_{\rm X-ray-emitting, <R_{\rm vir}} = 0.09 \pm 0.02$), respectively. If considering also the side effect in temperature, 0.1 dex and 0.2 dex uncertainties yield $M_{\rm X-ray-emitting, <R_{\rm vir}}=1.5\pm0.7\times10^{11}M_\odot$ ($f_{\rm X-ray-emitting, <R_{\rm vir}} = 0.11 \pm 0.05$) and $M_{\rm X-ray-emitting, <R_{\rm vir}}=2.0\pm1.2\times10^{11}M_\odot$ ($f_{\rm X-ray-emitting, <R_{\rm vir}} = 0.15 \pm 0.09$), respectively. In summary, for MW-mass galaxies, if no spectral constraints on the X-ray-emitting gas are available and the temperature is instead inferred from the halo mass, an overestimation (underestimation) of the halo mass by $0.1\,\rm dex$ can lead to an underestimation (overestimation) of $f_{\rm X-ray-emitting, <R_{\rm vir}}$ by about 30\% (60\%).

\subsection{Metallicity dependence} \label{Sec_Z}
Measuring the metallicity of the CGM is observationally challenging. 
From pulsar dispersion measure toward the Large Magellanic Cloud, the metallicity is constrained to be $Z \gtrsim 0.3\,Z_\odot$ \citep{Anderson2010,Miller2015}.
Using integrated soft X-ray emission observed by eROSITA in the eFEDS field, \citet{Ponti2023} obtain $Z \sim 0.1\,Z_\odot$ for the outer CGM of the MW. Within $<50\,\rm kpc$, the CGM metallicity of nearby massive galaxies ($M_* \approx 3\times 10^{11}\,M_\odot$) has been measured to be $\sim0.1-0.2\,Z_\odot$ \citep{Bogdan2013,Anderson2016,Bogdan2017}. 
For massive galaxy clusters, the metallicity of the ICM is $\sim$0.3$\,Z_\odot$ beyond $0.3\,R_{\rm 500c}$, rising toward the central galaxy to $0.6-1\,Z_\odot$ \citep{Mernier2018, Gastaldello2021,Sarkar2022}. 

We test several metallicity assumptions: i) a constant $Z = 0.1\,Z_\odot$, ii) a constant $Z = Z_\odot$ (used as an upper limit, though not physically realistic), and iii) a radially varying profile inspired by ICM measurements, with $Z = 0.3\,Z_\odot$ beyond $0.3\,R_{\rm 500c}$ (44 kpc) and increasing inwards as $Z = 0.2\,Z_\odot (r/R_{\rm 500c})^{-0.3}$, reaching $0.8\,Z_\odot$ at $0.01\,R_{\rm 500c}$. The inferred $f_{\rm X-ray-emitting, <R_{\rm vir}}$ values are shown in Fig.~\ref{Fig_fb}. In general, $f_{\rm X-ray-emitting, <R_{\rm vir}}$ varies between $0.05-0.16$. The accurate measurement of the X-ray-emitting CGM metallicity relies on the X-ray microcalorimeter telescopes \citep{Barret2023,XRISM2020,ZhangYN2022}.

\subsection{Other assumptions} \label{Sec_other}

There are additional systematic uncertainties in estimating the X-ray-emitting gas mass from the observed stacked 0.5--2 keV $S_{\rm X}$ profile that are not covered in the above discussion. 
We assumed that the mean $S_{\rm X}$ profile of MW-mass galaxies measured by stacking 30,825 galaxies with $\log(M_*/M_\odot)=10.5-11.0$ represents the $S_{\rm X}$ profile of a galaxy with $M_{\rm vir}=1.3\times10^{12}\,\rm M_\odot$ \citep{Zhang2024profile}. 
However, two sources of scatter in the stacking measurements must be taken into account.
First, there is a large spread ($\sigma_{\log(M_{\rm h}),16-84\%}=$0.8 dex) in the halo mass distribution of the stacked galaxy sample. The stacked signal is the mean value from all galaxies, while the more massive and brighter galaxies contribute more to the X-ray signal. As a result, the mean $S_{\rm X}$ measured in the stacking is not the median $S_{\rm X}$ of the stacked galaxy sample. 
Second, there is intrinsic scatter in $S_{\rm X}$ among galaxies of similar halo mass.
While the discussion in Sect.~\ref{Sec_Mh} provide a preliminary estimation of the impact of halo mass distribution, a forward modeling that accounts for the halo mass and temperature distributions of the X-ray–emitting gas, as well as the intrinsic scatter as a function of halo mass over the range from $10^{12}M_\odot$ to $10^{14}M_\odot$ is necessary to explain the observed $S_{\rm X}$ profile and scaling relations \citep[e.g.,][]{Comparat2025}.

While plasma in collisionally ionised equilibrium is usually assumed to explain the X-ray emission from the hot CGM, several additional physical processes, though likely subdominant, may also contribute. To increase the completeness of the discussion, we briefly list these processes here without quantifying their impact:
\begin{itemize}
    \item The charge exchange occurring at the interfaces between neutral and ionized gas (e.g., between cooling inflows and volume-filling hot gas) may contribute to the soft X-ray emission \citep[e.g.,][]{Wang2012_CX,Lopez2020}.
    \item Near the galaxy disk and in the CGM outskirts where gas densities are $\sim 10^{-5}\,\rm cm^{-3}$, photoionization can modify the plasma ionization balance and the X-ray spectrum \citep[e.g.,][]{Fox2005, Qu2018}.
    \item The resonant scattering process can modify the $S_{\rm X}$ profile slope, especially in the inner halo \citep{Nelson2023}.
    \item The inverse Compton scattering between cosmic rays and the cosmic microwave background (CMB) emits X-rays that may contribute to the total signal around MW-mass galaxies \citep{Ji2020,Stein2023,Silich2025, Hopkins2025,Lu2025}. X-ray spectral analysis (especially emission and absorption line features) will be critical to disentangling thermal and non-thermal emission components.
\end{itemize}

In this work, we have examined the effects of varying $n_{\rm H}$, $T$, $Z$, and halo mass assumptions individually. But in reality, the assumptions are not independent of each other. For instance, the assumed halo mass directly influences the expected temperature, and the physical properties of the CGM are jointly determined by theoretical assumptions, feedback models, and baryonic processes, as treated self-consistently in frameworks such as ExpCGM and in cosmological simulations.
Propagating the combined uncertainties shown in Fig.~\ref{Fig_fb} into a total error is therefore non-trivial and will require dedicated future analysis.

\section{Conclusions}

We show that the inferred X-ray-emitting baryon mass in the CGM of MW-mass galaxies derived from the mean $0.5-2$ keV energy band $S_{\rm X}$ profile measured in \citet{Zhang2024profile} is highly sensitive to assumptions about gas temperature, halo mass, metallicity, and gas density profile. 
Under a fiducial model with an isothermal $T = 0.12\,\mathrm{keV}$, $Z = 0.3Z_\odot$, and a beta profile with $\beta = 0.4$, we obtain $f_{\rm X-ray-emitting, <R_{\rm vir}} = 0.084 \pm 0.005$. Modifying the gas density slope to $\beta = 0.5$ or assuming a modified Vikhlinin profile with a steep $n_{\rm H}\sim r^{-3.2}$ at large radii ($>200\,\rm kpc$) changes $f_{\rm X-ray-emitting, <R_{\rm vir}}$ by 20--30\%. When varying temperature assumptions, including isothermal temperatures (0.17 keV, 0.08 keV, and temperatures extrapolated from the $T-M_{\rm 500c}$ scaling relation observed in galaxy clusters), and considering the log-normal temperature distribution, $f_{\rm X-ray-emitting, <R_{\rm vir}}$ can vary by more than a factor of four. 
Adopting a physically motivated ExpCGM model yields $f_{\rm X-ray-emitting, <R_{\rm vir}} = 0.113 \pm 0.011$, and comparison with the LC-TNG simulations gives $f_{\rm X-ray-emitting, <R_{\rm vir}} \approx 0.047 \pm 0.007$. A 0.2 dex uncertainty in halo mass further introduces up to a factor of four variation in $f_{\rm X-ray-emitting, <R_{\rm vir}}$. Varying the metallicity from $0.1Z_\odot$ to $Z_\odot$ changes $f_{\rm X-ray-emitting, <R_{\rm vir}}$ by up to a factor of three. 

In addition, other emission mechanisms and scatter in the halo mass further compound the uncertainties in $f_{\rm X-ray-emitting, <R_{\rm vir}}$. 
As such, the $0.5-2$ keV energy band X-ray surface brightness profile alone is insufficient to robustly constrain the baryon mass of the CGM for MW-mass galaxies. A differentiable forward-modeling approach, guided by simulations and incorporating realistic systematics, to synthesize the information from SZ, FRBs, and X-rays, is needed. Joint modeling of X-ray and SZ observations with halo masses calibrated by weak lensing has narrowed down the uncertainty of the gas mass \citep[e.g.,][]{Oren2024, Siegel2025}. 
X-ray spectral measurements with eROSITA's CCD spectral resolution can put constraints on the CGM temperature \citep{Toptun2025}. Ultimately, high-resolution spectral analysis enabled by future X-ray telescopes (e.g., New{\it Athena}, HUBS, Arcus, LEM) will be essential to break the degeneracies discussed here and resolve the `missing' baryon problem in galaxies \citep{Arcus2016, Cui2020,LEM2022, ZhangYN2022,Barret2023,Cruise2025,ZhangYN2025}.  

\begin{acknowledgements}

We thank Rahul Ramesh and Xiaoyuan Zhang for the helpful discussions. We thank Philip F. Hopkins for noticing the virial temperature error in Zhang et al. 2024, which motivated this paper. 
This project acknowledges financial support from the European Research Council (ERC) under the European Union's Horizon 2020 research and innovation program HotMilk (grant agreement No. 865637). 
GP acknowledges support from the European Research Council (ERC) under the European Union's Horizon 2020 research and innovation program HotMilk (grant agreement No. 865637), support from Bando per il Finanziamento della Ricerca Fondamentale 2022 dell'Istituto Nazionale di Astrofisica (INAF): GO Large program and from the Framework per l'Attrazione e il Rafforzamento delle Eccellenze (FARE) per la ricerca in Italia (R20L5S39T9). 
T. F. is supported by the National Natural Science Foundation of China under Nos. 11890692, 12133008, 12221003. We acknowledge the science research grants from the China Manned Space Project with No. CMS-CSST-2021-A04 and No. CMS-CSST-2025-A10.
\\

\end{acknowledgements}

\bibliographystyle{aa}
\bibliography{ref.bib}

\begin{appendix} 

\section{Detected baryon mass of MW-mass galaxies}\label{Sec_MWb}

The X-ray-emitting or X-ray-absorbing gas mass measured for MW or MW-mass galaxies spans a wide range across studies: $>6.1\times10^{10}\,\rm M_\odot$ within 139 kpc \citep[][also see the discussion in \citet{Wang2012} and \citet{Gupta2017}]{Gupta2012}, $3.3-9.8\times10^{10}\,\rm M_\odot$ within 200 kpc \citep[][]{Miller2013}, $0.2-1.3\times10^{11}\,\rm M_\odot$ within 250 kpc \citep{Nicastro2016}, $2.8-5.2\times10^{10}\,\rm M_\odot$ within 250 kpc \citep[][]{Fang2015,Hodges-Kluck2016,Bregman2018,Miller2015,LiYY2017}, $9.4\times10^{10}\,\rm M_\odot$ \citep[][]{Faerman2017}, and $1-1.7\times10^{11}\,\rm M_\odot$ within 200 kpc \citep[][]{Nicastro2023}. The main sources of uncertainty in these measurements include the metallicity assumption, deprojection of the gas density profile, and contamination from the solar wind charge exchange \citep{Bregman2018,Qu2022}. By stacking 12 galaxies with $\log(M_*/M_\odot)=10.6-11.0$, the thermal SZ signal is detected out to $\sim 250\,\rm kpc$ with $\approx4\sigma$ significance, which suggests a gas mass $=1.2-2.6\times10^{11}\,M_\odot$ within 250 kpc, by assuming the thermal SZ signal arises from a $T=0.26\,\rm keV$ plasma \citep{Bregman2022}.

\begin{figure}[ht]
\centering
\includegraphics[width=0.99\columnwidth]{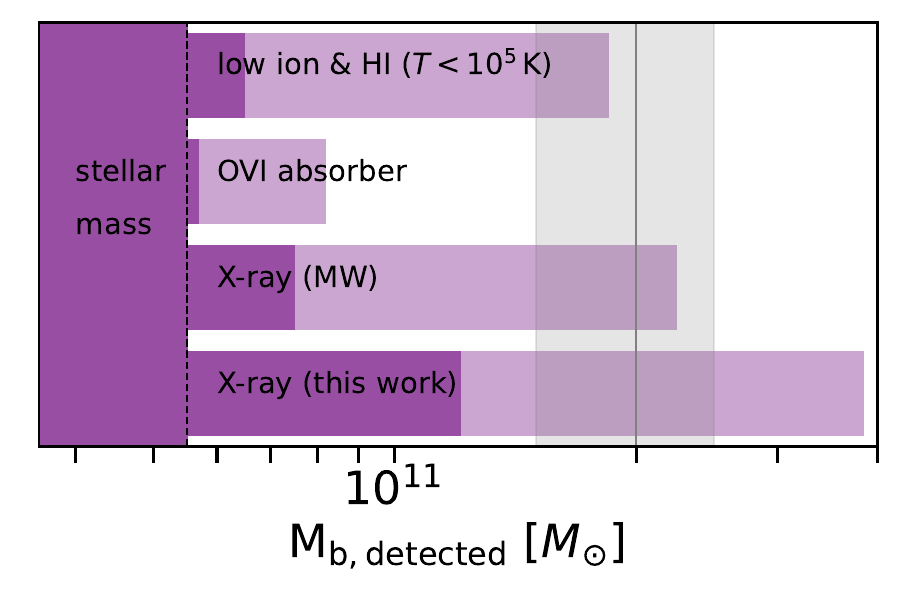}
\caption{Baryon mass detected through different tracers as summarized in Table~\ref{Table_MWb} and Appendix~\ref{Sec_MWb}. The baryon masses are measured for the MW or $L_*$ galaxies (noting that the definition of $L_*$ varies across studies, spanning a broad range in stellar and halo mass). For reference, we take a stellar mass of $5.5\times10^{10}\,M_\odot$. Dark and light purple bars indicate the lower and upper limits of detected baryon mass, respectively. The upper limit derived in this work excludes the case of an isothermal $T=0.08\,\rm keV$. The gray line (and shaded bar) denotes the expected baryon mass of galaxies with $M_{\rm h}=1.3\times10^{12}\,M_\odot$ ($1-1.6\times10^{12}\,M_\odot$).}
\label{Fig_MWb}
\end{figure}

The mass of warm ($T=10^5-10^6\,\rm K$) CGM traced by UV absorption lines (e.g., O VI) spans $2\times10^9-3\times10^{10}\,\rm M_\odot$, where the large uncertainty primarily originates from the assumption of ionization fraction \citep{Tumlinson2011,Peeples2014,TumlinsonPeeplesWerk_2017ARAA..55..389T, ChenHW2024}.
Using some low-ionization (e.g., Si II, Si III) absorbers in the COS-Halos survey and stacked H$\alpha$ emission, the cool ($T<10^5\,\rm K$) gas mass is estimated to be $M_{\rm CGM}=4\times10^{10}-1.3\times10^{11}M_\odot$ within $R_{\rm vir}$ \citep{Werk2014,Prochaska2017,ZhangHN2018a}, where metallicity uncertainties have been explicitly considered. An analytic model developed by \citet{Faerman2023} to interpret the COS-Halos measurements suggests a smaller cool gas mass of $\approx10^{10}\,M_\odot$. A summary of the baryon mass measurements for different CGM phases, derived from various observational methods, is presented in Fig.~\ref{Fig_MWb} and Table~\ref{Table_MWb}.

\begin{sidewaystable}[]
    \centering
    \caption{Baryon budget in the CGM of MW-mass galaxies}
    \label{Table_MWb}
    \begin{tabular}{ccccccc}
\hline \hline
Reference & Baryon mass [$M_\odot$] & Largest& Targeted galaxy & Note \\
&&radius [kpc]&&\\
\hline
\citet{Gupta2012}  &$>6.1\times10^{10}$ &139&MW$^*$&O VII absorber, Z=0.3$Z_\odot$ \\
\citet{Miller2013}  &$3.3-9.8\times10^{10}$ &200&MW$^*$&O VII absorber, Z=$Z_\odot$ \\
\citet{Miller2015}  &$3.5-5.2\times10^{10}$ &250&MW$^*$&O VIII emission, Z=0.3$Z_\odot$\\
\citet{Nicastro2016}  &$0.2-1.3\times10^{11} $&250&MW$^*$&O VII absorbers, Z=0.3$Z_\odot$ \\
\citet{LiYY2017}  &$3-4\times10^{10}$ &250&MW$^*$&O VIII emission, Z=0.3$Z_\odot$\\
\citet{Faerman2017}  &$9.4\times10^{10}$ &250&MW$^*$&X-ray absorption and emission, Z=0.5$Z_\odot$\\
\citet{Bregman2022}&$1.2-2.6\times10^{11}$&250&$M_{\rm h}\approx 10^{12}\,\rm M_\odot$&SZ stacking, T=0.26 keV\\
\citet{Nicastro2023}  &$1-1.7\times10^{11}$ &200&$\log(M_{\rm h})=11.9-12.4$&O VII absorbers, Z=0.3$Z_\odot$\\
\hline
\citet{Tumlinson2011}&$0.2-2.0\times10^{10}$&150&$\log(M_{\rm h}/M_\odot)=11.6-12.1$&O VI absorber, $f_{\rm O VI}$=0.2\\
\citet{Peeples2014}&$1.1-2.7\times10^{10}$&150&$\log(M_*/M_\odot)=9.3-11.6$&O VI absorber, Z=0.3$Z_\odot$\\
\hline
\citet{Werk2014} & $0.7-1.2\times10^{11}$&160& $
\log(M_*/M_\odot)_{\rm med}=10.5$ & low-ion UV absorbers, COS-Halos\\
\citet{Prochaska2017} & $0.5-1.3\times10^{11}$&160& $0.3 < L/L_* < 2$ & Lyman limit, COS-Halos\\
\citet{ZhangHN2020} & $4.1-7.4\times10^{10}$&50(projected)& $M_{\rm h}\approx1.6\times10^{12}\,\rm M_\odot$ & H$\alpha$ emission stacking, SDSS\\
\hline

\end{tabular}
\tablefoot{Baryon mass (second column) within a certain radius (third column) measured in the work (first column) for our MW (denoted as MW$^*$) or $L_*$ galaxies defined in the respective studies (fourth column).
The final column summarizes the detection method and key assumptions adopted in each work.
}
\end{sidewaystable}

\section{Discussions about more massive galaxies} \label{Sec_moremassive}

In the main text, we discussed how the derived $f_{\rm X-ray-emitting, <R_{\rm vir}}$ from the observed $S_{\rm X}$ profile depends on assumptions about the gas density profile, temperature, halo mass uncertainty, and metallicity for MW-mass galaxies. The mean $S_{\rm X}$ profiles of galaxies in stellar mass bins $\log(M_*/M_\odot)=11.0-11.25$ (the `M31' bin), $\log(M_*/M_\odot)=11.25-11.5$ (the `2M31' bin), as well as in halo mass bins $\log(M_{\rm 200m}/M_\odot)=12.5-13.0$, $\log(M_{\rm 200m}/M_\odot)=13.0-13.5$, and $\log(M_{\rm 200m}/M_\odot)=13.5-14.0$ are measured in \citet{Zhang2024profile}. Here, we extend the discussion in the main text to these five higher-mass bins to quantify how the derived $f_{\rm X-ray-emitting, <R_{\rm vir}}$ depends on various assumptions. 

For each of the five mass bins, we take the fiducial assumption of a beta-model gas density profile (Eq.~\ref{Eq_beta}), an isothermal temperature of $T_{\rm 500c}$ (Eq.~\ref{Eq_Tphi}), a metallicity of $0.3Z_\odot$, and the median $M_{\rm vir}$, $M_{\rm 500c}$, $R_{\rm vir}$, and $R_{\rm 500c}$ for galaxies in that bin. 
Unlike in the main-text analysis, we treat $\beta$ and $r_{\rm c}$ as free parameters when fitting to the observed $S_{\rm X}$ profiles.
We then vary the fiducial assumptions individually as follows: i) a modified Vikhlinin profile (Eq.\ref{Eq_Vikh}); ii) isothermal temperatures extrapolated from the $T-M_{\rm 500c}$ scaling relation assuming slopes of 0.57 and 0.66 (Sect.~\ref{Sec_T}); iii) a log-normal temperature distribution with a median temperature equal to the fiducial value and scatter $\sigma=0.3$; iv) the ExpCGM framework (Sect.~\ref{Sec_Tlog}); v) considering halo mass uncertainties of 0.1 dex and 0.2 dex (Sect.~\ref{Sec_Mh}); and vi) considering different metallicities (Sect.~\ref{Sec_Z}).

\begin{sidewaysfigure}[ht]
\centering
    \includegraphics[width=0.75\linewidth]{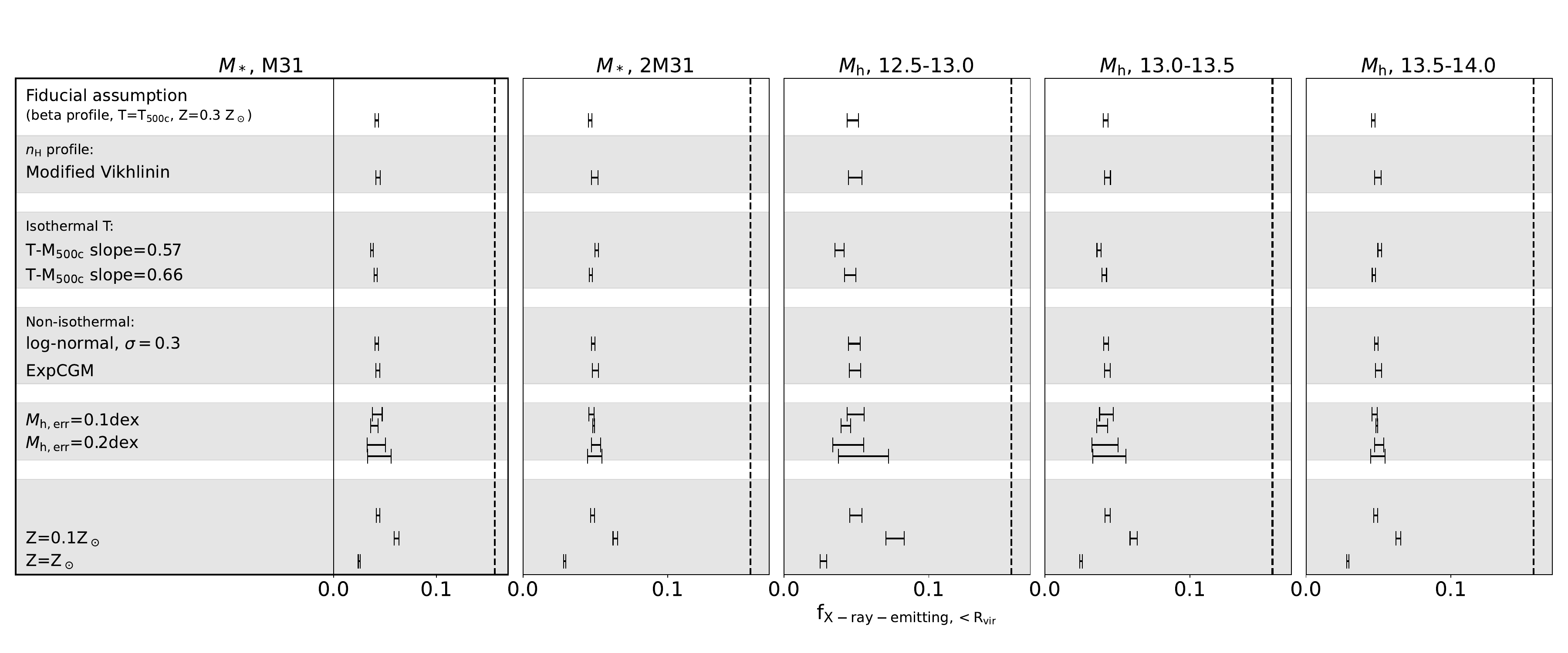}
\caption{$f_{\rm X-ray-emitting, <R_{\rm vir}}$ values derived from the $S_{\rm X}$ profiles measured by \citet{Zhang2024profile}, under the different assumptions listed in Table~\ref{Table_fbMs} and Table~\ref{Table_fbMh}. Results are shown for galaxies in stellar mass bins $\log(M_*/M_\odot)=11.0-11.25$ (M31), $\log(M_*/M_\odot)=11.25-11.5$ (2M31), as well as for halo mass bins $\log(M_{\rm 200m}/M_\odot)=12.5-13.0$, $\log(M_{\rm 200m}/M_\odot)=13.0-13.5$, and $\log(M_{\rm 200m}/M_\odot)=13.5-14.0$. The vertical dashed line indicates the cosmological baryon fraction ($f_{\rm b}=$0.157).}
\label{Fig_fb_broad}
\end{sidewaysfigure}

We list in Table~\ref{Table_fbMs} and Table~\ref{Table_fbMh} the adopted assumptions and the resulting $f_{\rm X-ray-emitting, <R_{\rm vir}}$. The results are also plotted in Fig.~\ref{Fig_fb_broad}. We find that for galaxies with higher stellar or halo mass, $f_{\rm X-ray-emitting, <R_{\rm vir}}$ becomes less sensitive to the assumed temperature, and its uncertainty correspondingly decreases.

\begin{table*}[]
    \centering
    \caption{Properties of X-ray-emitting gas under different assumptions.}
    \label{Table_fbMs}
    \begin{tabular}{ccccccccccc}
\hline \hline
 &&$\log(M_*/M_\odot)$& $\log(M_*/M_\odot)$ & $\log(M_*/M_\odot)$ \\
 &&$10.5-11.0$ (MW)& $11.0-11.25$ (M31) & $11.25-11.5$ (2M31)\\
\hline 
Fiducial&T [keV]&0.12& 0.18&0.36\\
(beta model)&$\beta$&0.4&$0.37^{+0.02}_{-0.03}$&$0.37\pm0.01$\\
($T=T_{\rm 500c}$)&$r_{c}$ [kpc]&5&$6^{+6}_{-4}$&$10^{+4}_{-3}$\\
(0.3$Z_\odot$)&$M_{\rm vir}$ [$M_\odot$]&$1.3\times10^{12}$&$4.2\times10^{12}$&$1.2\times10^{13}$ \\
&$M_{\rm 500c}$ [$M_\odot$]&$8.7\times10^{11}$&$2.7\times10^{12}$&$7.6\times10^{12}$\\
&$R_{\rm vir}$ [kpc]&285&396&551\\
&$f_{\rm X-ray-emitting, <R_{\rm vir}}$&$0.097\pm0.007$&$0.057\pm0.004$&$0.054\pm0.002$\\
\hline
Modified Vikhlinin&$\beta$&0.4&$0.36^{+0.04}_{-0.14}$&$0.35^{+0.02}_{-0.10}$\\
(Eq.~\ref{Eq_Vikh})&$r_c$ [kpc]&5&$7^{+6}_{-3}$&$9^{+4}_{-3}$\\
&$\epsilon$&4&-&$2.5_{-2.0}$\\
&$r_{\rm s}$ [kpc]&200&$\infty$&$930_{-900}$\\
&$f_{\rm X-ray-emitting, <R_{\rm vir}}$&$0.078\pm0.005$&$0.059\pm0.005$&$0.056\pm0.002$\\
\hline
$T \propto M_{\rm 500c}^{0.57}$&T [keV]& 0.15&0.29&0.53\\
&$f_{\rm X-ray-emitting, <R_{\rm vir}}$&$0.068\pm0.004$&$0.044\pm0.003$&$0.046\pm0.001$\\
\hline
$T \propto M_{\rm 500c}^{0.66}$&T [keV]& 0.09&0.19&0.38\\
&$f_{\rm X-ray-emitting, <R_{\rm vir}}$&$0.19\pm0.01$&$0.054\pm0.004$&$0.053\pm0.001$\\
\hline
Log-normal T& $T_{\rm mean}$ [keV]&0.125& 0.19&0.37\\
&$f_{\rm X-ray-emitting, <R_{\rm vir}}$&$0.086\pm0.006$&$0.057\pm0.004$&$0.053\pm0.001$\\
\hline
ExpCGM&$\alpha_0$&$0.6\pm0.2$&$0.8\pm0.1$&$0.81\pm0.04$\\
&$\alpha_1$&$1.1\pm0.3$&$0.9^{+0.2}_{-0.1}$&$0.9\pm0.1$\\
&$f_{\rm X-ray-emitting, <R_{\rm vir}}$&$0.11\pm0.01$&$0.058\pm0.004$&$0.054\pm0.001$\\
\hline
$\log(M_{\rm vir})+0.1$ dex&T [keV]&Fiducial&Fiducial&Fiducial\\
&$f_{\rm X-ray-emitting, <R_{\rm vir}}$&$0.083\pm0.005$&$0.047\pm0.003$&$0.046\pm0.001$\\
&$T \propto M_{\rm 500c}^{0.62}$&0.13&0.27&0.51\\
&$f_{\rm X-ray-emitting, <R_{\rm vir}}$&$0.069\pm0.005$&$0.042\pm0.003$&$0.043\pm0.001$\\
\hline
$\log(M_{\rm vir})-0.1$ dex&T [keV]&Fiducial&Fiducial&Fiducial\\
&$f_{\rm X-ray-emitting, <R_{\rm vir}}$&$0.100\pm0.006$&$0.072\pm0.005$&$0.063\pm0.002$\\
&$T \propto M_{\rm 500c}^{0.62}$&0.10&0.2&0.38\\
&$f_{\rm X-ray-emitting, <R_{\rm vir}}$&$0.16\pm0.01$&$0.057\pm0.004$&$0.057\pm0.002$\\
\hline
$\log(M_{\rm vir})+0.2$ dex&T [keV]&Fiducial&Fiducial&Fiducial\\
&$f_{\rm X-ray-emitting, <R_{\rm vir}}$&$0.074\pm0.005$&$0.040\pm0.003$&$0.040\pm0.001$\\
&$T \propto M_{\rm 500c}^{0.62}$&0.15&0.31&0.58\\
&$f_{\rm X-ray-emitting, <R_{\rm vir}}$&$0.051\pm0.003$&$0.037\pm0.003$&$0.038\pm0.001$\\
\hline
$\log(M_{\rm vir})-0.2$ dex&T [keV]&Fiducial&Fiducial&Fiducial\\
&$f_{\rm X-ray-emitting, <R_{\rm vir}}$&$0.109\pm0.007$&$0.097\pm0.006$&$0.073\pm0.002$\\
&$T \propto M_{\rm 500c}^{0.62}$&0.09&0.18&0.33\\
&$f_{\rm X-ray-emitting, <R_{\rm vir}}$&$0.255\pm0.017$&$0.069\pm0.004$&$0.066\pm0.002$\\
\hline
Z(R)&$f_{\rm X-ray-emitting, <R_{\rm vir}}$&$0.088\pm0.006$&$0.059\pm0.004$&$0.056\pm0.002$\\
0.1$Z_\odot$&$f_{\rm X-ray-emitting, <R_{\rm vir}}$&$0.16\pm0.01$&$0.093\pm0.006$&$0.080\pm0.002$\\
$Z_\odot$&$f_{\rm X-ray-emitting, <R_{\rm vir}}$&$0.054\pm0.004$&$0.032\pm0.002$&$0.032\pm0.001$\\
\end{tabular}
\tablefoot{We list the values of temperature, gas density profile, and $f_{\rm X-ray-emitting, <R_{\rm vir}}$ under the different assumptions described in the main text and Appendix.~\ref{Sec_moremassive}, for three stellar mass bins.
}
\end{table*}

\begin{table*}[]
    \centering
    \caption{Properties of X-ray-emitting gas under different assumptions.}
    \label{Table_fbMh}
    \begin{tabular}{ccccccccccc}
\hline \hline
 && $\log(M_{\rm 200m}/M_\odot)$& $\log(M_{\rm 200m}/M_\odot)$&$\log(M_{\rm 200m}/M_\odot)$ \\
 && $12.5-13.0$ & $13.0-13.5$&$13.5-14.0$ \\
\hline 
Fiducial&T [keV]&0.2&0.43&0.82\\
&$\beta$&$0.39\pm0.03$&$0.39\pm0.02$&$0.38\pm0.01$\\
&$r_{c}$ [kpc]&$7_{-4}^{+6}$&$11\pm3$&$16\pm3$\\
&$M_{\rm vir}$ [$M_\odot$]&$5\times10^{12}$&$1.6\times10^{13}$&$4.3\times10^{13}$ \\
&$M_{\rm 500c}$ [$M_\odot$]&$3.2\times10^{12}$&$1.0\times10^{13}$&$2.6\times10^{13}$\\
&$R_{\rm vir}$ [kpc]&419&611&839\\
&$R_{\rm 500c}$ [kpc]&221&320&437\\
&$f_{\rm X-ray-emitting, <R_{\rm vir}}$&$0.048\pm0.004$&$0.042\pm0.002$&$0.046\pm0.001$\\
&$f_{\rm X-ray-emitting, <R_{\rm 500c}}$&$0.023\pm0.002$&$0.020\pm0.001$&$0.022\pm0.001$\\
\hline
Modified Vikhlinin&$\beta$&$0.38^{+0.04}_{-0.09}$&$0.38^{+0.02}_{-0.04}$&$0.36\pm0.03$\\
&$r_c$ [kpc]&$7_{-4}^{+7}$&$10\pm4$&$13_{-4}^{+5}$\\
&$\epsilon$&-&-&-\\
&$r_{\rm s}$ [kpc]&$\infty$&$\infty$&$\infty$\\
&$f_{\rm X-ray-emitting, <R_{\rm vir}}$&$0.049\pm.005$&$0.043\pm0.002$&$0.050\pm0.002$\\
&$f_{\rm X-ray-emitting, <R_{\rm 500c}}$&$0.024\pm0.002$&$0.021\pm0.001$&$0.024\pm0.001$\\
\hline
$T \propto M_{\rm 500c}^{0.57}$&T [keV]& 0.33&0.62&1.1\\
&$f_{\rm X-ray-emitting, <R_{\rm vir}}$&$0.038\pm0.003$&$0.037\pm0.001$&$0.051\pm0.001$\\
&$f_{\rm X-ray-emitting, <R_{\rm 500c}}$&$0.019\pm0.001$&$0.018\pm0.001$&$0.025\pm0.001$\\
\hline
$T \propto M_{\rm 500c}^{0.66}$&T [keV]& 0.22&0.46&0.86\\
&$f_{\rm X-ray-emitting, <R_{\rm vir}}$&$0.046\pm0.004$&$0.041\pm0.002$&$0.047\pm0.001$\\
&$f_{\rm X-ray-emitting, <R_{\rm 500c}}$&$0.022\pm0.002$&$0.020\pm0.001$&$0.023\pm0.001$\\
\hline
Log-normal T& $T_{\rm mean}$ [keV]&0.21&0.45&0.86\\
&$f_{\rm X-ray-emitting, <R_{\rm vir}}$&$0.049\pm0.004$&$0.042\pm0.002$&$0.049\pm0.001$\\
&$f_{\rm X-ray-emitting, <R_{\rm 500c}}$&$0.023\pm0.002$&$0.020\pm0.001$&$0.024\pm0.001$\\
\hline
ExpCGM&$\alpha_0$&$0.8\pm0.1$&$0.79\pm0.04$&$0.66\pm0.02$\\
&$\alpha_1$&$0.9\pm0.2$&$1.1\pm0.1$&$1.7\pm0.1$\\
&$f_{\rm X-ray-emitting, <R_{\rm vir}}$&$0.049\pm0.004$&$0.043\pm 0.002$&$0.050\pm0.002$\\
\hline
$\log(M_{\rm vir})+0.1$ dex&T [keV]&Fiducial&Fiducial&Fiducial\\
&$f_{\rm X-ray-emitting, <R_{\rm vir}}$&$0.040\pm0.003$&$0.036\pm0.002$&$0.044\pm0.001$\\
&$f_{\rm X-ray-emitting, <R_{\rm 500c}}$&$0.019\pm0.001$&$0.018\pm0.001$&$0.022\pm0.001$\\
&$T \propto M_{\rm 500c}^{0.62}$&0.3&0.6&1.09\\
&$f_{\rm X-ray-emitting, <R_{\rm vir}}$&$0.036\pm0.003$&$0.034\pm0.001$&$0.047\pm0.001$\\
&$f_{\rm X-ray-emitting, <R_{\rm 500c}}$&$0.017\pm0.001$&$0.017\pm0.001$&$0.023\pm0.001$\\
\hline
$\log(M_{\rm vir})-0.1$ dex&T [keV]&Fiducial&Fiducial&Fiducial\\
&$f_{\rm X-ray-emitting, <R_{\rm vir}}$&$0.059\pm0.005$&$0.049\pm0.002$&$0.050\pm0.001$\\
&$f_{\rm X-ray-emitting, <R_{\rm 500c}}$&$0.028\pm0.002$&$0.024\pm0.001$&$0.024\pm0.001$\\
&$T \propto M_{\rm 500c}^{0.62}$&0.23&0.46&0.83\\
&$f_{\rm X-ray-emitting, <R_{\rm vir}}$&$0.049\pm0.004$&$0.045\pm0.002$&$0.051\pm0.001$\\
&$f_{\rm X-ray-emitting, <R_{\rm 500c}}$&$0.023\pm0.002$&$0.022\pm0.001$&$0.025\pm0.001$\\
\hline
$\log(M_{\rm vir})+0.2$ dex&T [keV]&Fiducial&Fiducial&Fiducial\\
&$f_{\rm X-ray-emitting, <R_{\rm vir}}$&$0.034\pm0.003$&$0.032\pm0.001$&$0.043\pm0.001$\\
&$f_{\rm X-ray-emitting, <R_{\rm 500c}}$&$0.017\pm0.001$&$0.016\pm0.001$&$0.021\pm0.001$\\
&$T \propto M_{\rm 500c}^{0.62}$&0.34&0.69&1.25\\
&$f_{\rm X-ray-emitting, <R_{\rm vir}}$&$0.031\pm0.003$&$0.031\pm0.001$&$0.046\pm0.001$\\
&$f_{\rm X-ray-emitting, <R_{\rm 500c}}$&$0.015\pm0.001$&$0.015\pm0.001$&$0.022\pm0.001$\\
\hline
$\log(M_{\rm vir})-0.2$ dex&T [keV]&Fiducial&Fiducial&Fiducial\\
&$f_{\rm X-ray-emitting, <R_{\rm vir}}$&$0.075\pm0.006$&$0.057\pm0.002$&$0.056\pm0.001$\\
&$f_{\rm X-ray-emitting, <R_{\rm 500c}}$&$0.036\pm0.002$&$0.028\pm0.001$&$0.027\pm0.001$\\
&$T \propto M_{\rm 500c}^{0.62}$&0.2&0.4&0.7\\
&$f_{\rm X-ray-emitting, <R_{\rm vir}}$&$0.058\pm0.004$&$0.052\pm0.002$&$0.055\pm0.001$\\
&$f_{\rm X-ray-emitting, <R_{\rm 500c}}$&$0.028\pm0.002$&$0.025\pm0.001$&$0.027\pm0.001$\\
\hline
Z(R)&$f_{\rm X-ray-emitting, <R_{\rm vir}}$&$0.050\pm0.004$&$0.043\pm0.002$&$0.048\pm0.001$\\
&$f_{\rm X-ray-emitting, <R_{\rm 500c}}$&$0.023\pm0.002$&$0.022\pm0.001$&$0.024\pm0.001$\\
0.1$Z_\odot$&$f_{\rm X-ray-emitting, <R_{\rm vir}}$&$0.077\pm0.006$&$0.061\pm0.002$&$0.064\pm0.002$\\
&$f_{\rm X-ray-emitting, <R_{\rm 500c}}$&$0.036\pm0.002$&$0.030\pm0.001$&$0.031\pm0.001$\\
$Z_\odot$&$f_{\rm X-ray-emitting, <R_{\rm vir}}$&$0.027\pm0.002$&$0.025\pm0.001$&$0.029\pm0.001$\\
&$f_{\rm X-ray-emitting, <R_{\rm 500c}}$&$0.013\pm0.001$&$0.012\pm0.001$&$0.014\pm0.001$\\
\end{tabular}
\tablefoot{Same as Table~\ref{Table_fbMs}, but for the three halo mass bins and including values of $f_{\rm X-ray-emitting, <R_{\rm 500c}}$ (ratio between X-ray-emitting gas mass within $R_{\rm 500c}$ and $M_{\rm 500c}$).
}
\end{table*}

\end{appendix}

\end{document}